\documentclass[aps,prb,twocolumn,superscriptaddress,showpacs,floatfix,10pt]{revtex4-1}

\usepackage{amsmath}
\usepackage{amsfonts}
\usepackage{graphicx}
\usepackage{color}
\usepackage{epstopdf}
\usepackage{natbib}
\usepackage{xfrac}
\usepackage{enumitem}
\usepackage[english]{babel}
\usepackage{booktabs}
\usepackage{soul}
\usepackage{xcolor}
\usepackage{lipsum}
\usepackage{tabularx}
\usepackage{multirow} 

\usepackage{float} 
\usepackage{subfigure}

\usepackage{booktabs}
\usepackage{multirow}
\usepackage{comment}

\begin{document}
\title{Perturbative Transfer Matrix Method for optical-pump-THz-probe of ultrafast dynamics}

\author{Yingshu~{Yang}}
\affiliation{School of Physical and Mathematical Sciences, Nanyang Technological University, Singapore, Singapore}

\author{Stefano~{Dal~Forno}}
\affiliation{School of Physical and Mathematical Sciences, Nanyang Technological University, Singapore, Singapore}

\author{Marco~{Battiato}}
\email{marco.battiato@ntu.edu.sg}
\affiliation{School of Physical and Mathematical Sciences, Nanyang Technological University, Singapore, Singapore}

\date{\today}

\begin{abstract}

Ultrafast optical excitations trigger in materials a range of dynamics. An interesting range of dynamics is the ones that unfold within the picosecond timescale, corresponding to the THz frequency range. Within this short timescale, the material's properties, and therefore its response to electromagnetic fields, dynamically change. For this reason, any THz radiation used to probe the material will interact with the multilayer undergoing time-dependent modifications of the dielectric responses within a single optical cycle (a very similar situation arises for THz radiation produced in spintronics THz emitters). Such interaction goes beyond typical quasistatic approaches. It is, therefore paramount to be able to describe accurately all the interaction of THz radiation with out-of-equilibrium multilayers. We develop here a theoretical framework called PTMM (Perturbative Transfer Matrix Method) to model the production of THz accurately and its interaction with multilayers undergoing ultrafast changes in their dielectric properties. 

That analysis allows us to propose two novel ways to utilize Terahertz time-domain spectroscopy. We will first show that a simple analysis of the time resolved Optical-Pump-Terahertz-Probe spectra can provide not only the time scale of processes that are slower than the THz pulse time-width, but it can accurately measure the timescale of sub-picosecond and faster processes as well. 
Further, we will apply our method to THz probed spintronics THz emitters, and compute the interference between the produced THz and the THz probe. We will show that an analysis of the delay-resolved transmission spectra allows for a direct measurement of the time distance between the laser excitation and the peak of the spin current, allowing an unprecedented insight into the ultrafast spin-to-charge conversion mechanism.

\end{abstract}


\maketitle

\section{Introduction}

Terahertz Time-Domain Spectroscopy (THz-TDS) emerges as a compelling experimental approach operating within the frequency spectrum of 0.1 THz to 10 THz. Its versatile application spans diverse domains.\cite{schmuttenmaer_exploring_2004} Particularly noteworthy is its pivotal role in the realm of materials science, where THz-TDS assumes significance in material characterisation and the determination of dielectric constants. Unlike conventional intensity-based measurements, THz-TDS directly captures transient electric field variations through its time-domain signal.\cite{schmuttenmaer_exploring_2004} Concomitantly, the integration of optically gated emission and detection methodologies within THz-TDS has given birth to an additional powerful technique: Optical Pump Terahertz Probe (OPTP). By harnessing the temporal behavior of THz pulses, this approach has gained substantial prominence in the exploration of non-equilibrium properties of materials. Specifically, it facilitates the investigation of phenomena such as charge carrier dynamics and transient alterations of mobilities.\cite{ulbricht_carrier_2011,george_ultrafast_2008,strait2011very} OPTP serves as an invaluable tool for comprehending material responses under dynamic circumstances, in contrast to the predominantly steady-state focus of traditional THz-TDS methodologies.

In OPTP experiments, the optical pump drives the system into an out-of-equilibrium state. Subsequently, a THz probe is sent onto the sample with variable delays, to investigate its properties. The out-of-equilibirum dynamics can be used to access specific information like hot carrier lifetime,\cite{afalla_photoconductivity_2019, gorodetsky_pump_2019} mobility,\cite{afalla_photoconductivity_2019, Hartono_effect_2019, krauspe_terahertz_2018} carrier density,\cite{zhang_direct_2019, mithun_dirac_2021} relaxation and decay mechanism,\cite{rao_ultrafast_2018, mithun_dirac_2021} as well as photoconductivity,\cite{afalla_photoconductivity_2019} which has to be deeply understood for the design and optimisation of new material based devices.\cite{lu_critical_2018}
The application of the OPTP experimental technique provides in-depth analysis of these processes as well as information on the non-equilibrium status of materials such as semiconductors, superconductors and metals, and other interesting materials such as, Mxenes,\cite{li_equilibrium_2018} perovskites,\cite{la-o-vorakiat_phonon_2017} MoS$_2$,\cite{kar_probing_2015} graphene,\cite{kar_tuning_2014, tomadin_ultrafast_2018, george_ultrafast_2008} topological insulators,\cite{ruan_terahertz_2021} which are all important component to computers, lasers, light-emitting devices, electrodes, information storage devices, and future electronic devices.\cite{ulbricht_carrier_2011, alberding_reduced_2016,kurner_towards_2014,koenig_wireless_2013,oftelie_towards_2020,takanashi_all_2020}

Interestingly, an optically-pumped multilayer can itself produce THz radiation. A famous example is the novel strategy to produce broadband THz radiation, the spintronic THz emitter (STE).\cite{kampfrath2013terahertz} After the optical excitation of a multilayer structure usually constructed using a ferromagnetic metal (FM) layer and a heavy metal (HM) layer, a rapid flow of spin current\cite{battiato_superdiffusive_2010,battiato_theory_2012} from the FM layer to the HM layer will happen perpendicularly to the sample plane and then be converted into a transversal charge current in the HM layer. This charge current in the HM layer will then produce THz.\cite{kampfrath2013terahertz} It means that in STE, a pump will both generate a time-dependent change in the dielectric response at THz frequencies (which can be probed with a THz probe) and produce a THz pulse.\cite{yang_modeling_2023,Liu_spintronic_2022,Agarwal_secondary_2023,agarwal_terahertz_2022,agarwal_interfacial_2023},

\begin{figure}[tb]
\centering
\includegraphics[width=\linewidth]{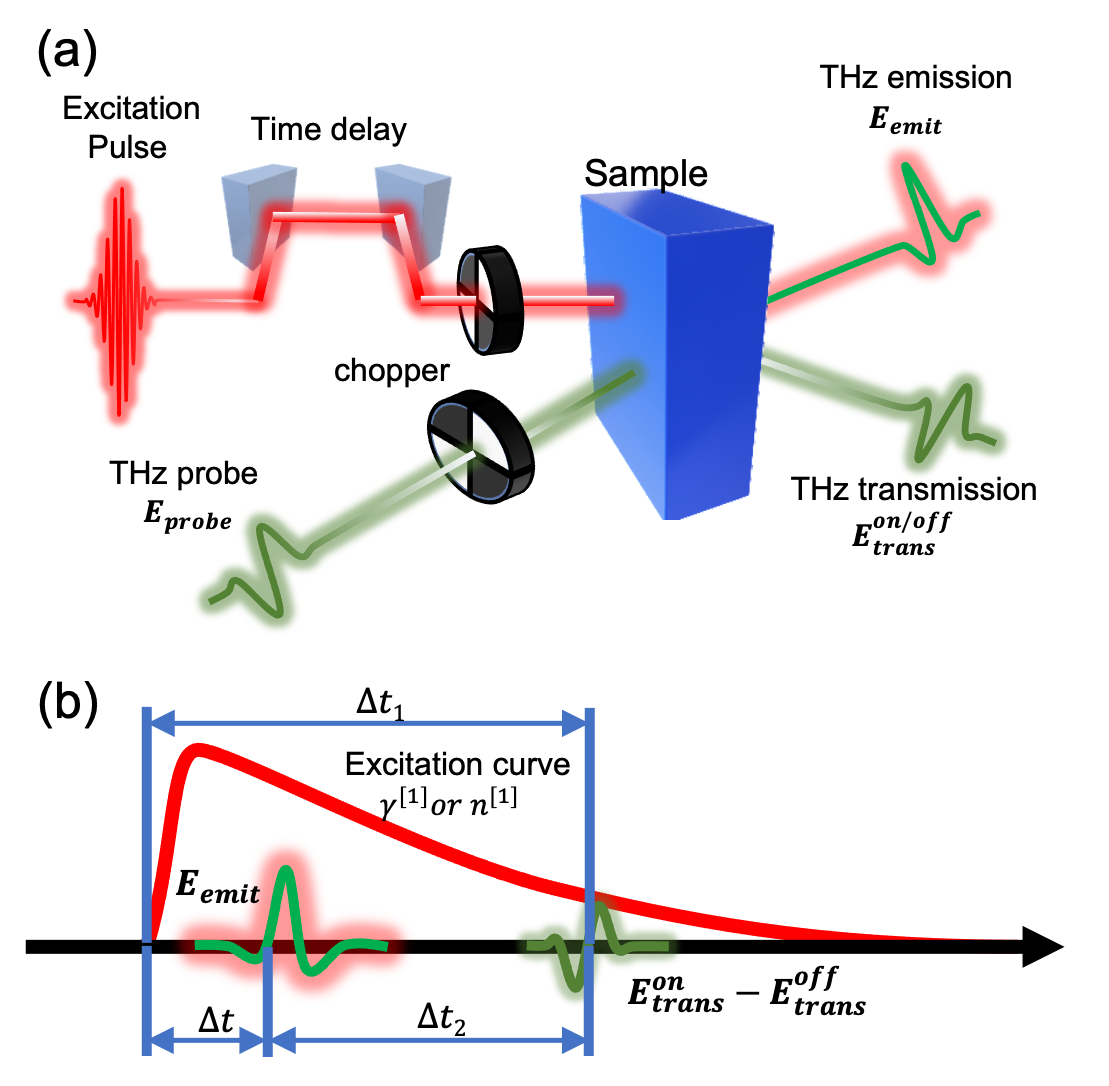}
\caption{Concept figure: A description of the OPTP and STE recipe. (a)
A THz probe pulse is sent into the system. The transmitted THz probe through an excited (pump-on, $E_{trans}^{on}$) and non-excited (pump-off, $E_{trans}^{off}$) are measured. A THz emission pulse ($E_{emit}$) can also be measured when the sample is a spintronic THz emitter. (b) A theoretical layout of the time delays can be accessed using our recipe.}
\label{fig:OPTP}
\end{figure}

Numerical and theoretical approaches have been developed to model simple THz-TDS\cite{yang_removal_2021} and OPTP experiments\cite{nemec_ultrafast_2005-1,nemec_ultrafast_2005,nemec_methodology_2002,kuzel_propagation_2007}, as well as the production of THz in STE,\cite{Yang_transfer_2021}. However, a comprehensive theoretical framework that can simultaneously address all these aspects of interaction of optically excited multilayers with THz radiation is still missing. 

In this manuscript, we will present a comprehensive approach that will allow to describe consistently the propagation and production of THz radiation within optically excited multilayers. The approach, the perturbative transfer matrix method (PTMM), is based on a perturbation expansion of the Maxwell-Drude system in the change of material properties compared to equilibrium. 

Theoretical analysis of the OPTP technique is usually performed under the assumption of quasistatic regime, which applies only when the  changes in the material properties of the multilayer happen over a timescale longer than the period of the used THz probe. This limits the application to the study of the de-excitation dynamics. Our PTMM is not limited to quasistatic regime and allows for an accurate description of the interaction of THz pulses with materials with properties changing in time in the sub-picosecond regime. \cite{beaurepaire_ultrafast_1996,eschenlohr_ultrafast_2013,carpene2008dynamics,kampfrath_ultrafast_2002,fann1992electron,sun1994femtosecond,suarez1995dynamics}
With our analysis we will be able to show that OPTP delay-resolved spectral maps contain a plethora of information: as an example, we will show a simple trick on how to extract the excitation time. Furthermore, being our PTMM fully compatible with the TMM with source\cite{Yang_transfer_2021}, we will be able to compute the effect of OPTP from spintronics THz emitters, where the spontaneously emitted THz pulse triggered by the optical pump, can overlap and interfere with a THz probe pulse. In this case, we will show how it is possible to use delay-resolved THz spectral maps to extract a number of characteristic features of the ultrafast spin-to-charge conversion phenomenon.

\section{Theoretical background}\label{Theoretical_background}

The Transfer Matrix Method (TMM) is often used to describe propagation of electromagnetic waves through multilayered systems, as it can provide accurate results by incorporating all left and right propagating waves into a single matrix.\cite{yang_removal_2021} The Transfer Matrix with source (TMMS), which includes a spatial source term for spintronic THz emission, has been developed and can be integrated into the basic TMM\cite{Yang_transfer_2021} to allow for the description of THz transmission and emission simultaneously. However, in OPTP experiments, the situation becomes more complex as an optical pump pulse is sent into the multilayer system before the THz probe. The optically excited sample undergoes a fast change of its dielectric responses, typically followed by a slower relaxation back to equilibrium.

A full description of such effects requires a number of steps. Assuming that the optical excitation still happens in the linear regime and that no saturation processes in absorption are to be accounted for, the deposition of optical photons' energy throughout the material can be obtained by using a standard TMM.\cite{beard_terahertz_2002,baxter_time_2007,yang_removal_2021} 

The next step in the description is how the deposited energy alters the dielectric responses of the different layers and what are the involved timescales. This is rather complex and will not be addressed in this work. We will assume that the frequency-dependent total permittivity $\epsilon_T [\omega]$ of the material can be written as the sum of a Drude component $\epsilon_D[\omega]$ and a remaining background component $\epsilon_B[\omega]$, which accounts for all the non-Drude contributions (e.g., the interband transitions) with a generic frequency dependence. We further assume that the background component of the permittivity is not modified by the laser. Conversely, we will allow the Drude parameters to be explicitly time-dependent. We assume that such time dependence is either known or is to be fitted to experimental results using the treatment that we will develop in the present work.

Assuming the time $t$ dependence of the Drude carrier density $n[t]$, and the inverse scattering lifetime $\gamma[t]$ known, the final step is to compute the propagation of a THz probe pulse through a multilayer in which one or more layers are undergoing that dynamic. Standard TMM cannot be used since a frequency-dependent total permittivity $\epsilon_T [\omega]$ cannot be written in such case. We then revert back to the Maxwell's equations. Before addressing the propagation through the entire multilayer, we first focus on a single layer at a time.

In the above situation, the electric $E$ and magnetic $H$ fields propagating along the $z$ axis are described by the Maxwell-Drude system:
\begin{equation} \label{eq_MaxwellDrude_main}
\begin{split}
	\partial_z E\left[z,\omega\right]=&-i\,\omega\,\mu[\omega]\,  H\left[z,\omega\right], \\
	\partial_z H\left[z,\omega\right]=&-i\,\omega\,\epsilon_{B}[\omega]\, E\left[z,\omega\right] + J[z,\omega], \\
	\partial_t J[z,t] =&-\gamma J[z,t] + \frac{n e^2}{m}E[z,t],
\end{split}
\end{equation}
where $\mu$ is the permeability of the medium, $J[z,t]$ is the current density (along the same direction as the electric field) induced by the Drude response,  $e$ is the electron charge, and $m$ the effective mass, $n$ is the number of carriers, and $\gamma$ the inverse scattering lifetime. In the case $n$ and $\gamma$ are constant in time, the last equation leads to the known expression for the Drude conductivity:$\sigma_D[\omega]=n e^2/(m\left(\gamma - i \omega \right)).$ The total permittivity $\epsilon_{T}[\omega]$ (i.e.~the background contribution and the Drude contribution) then reads $\epsilon_T [\omega]=\epsilon_{B}[\omega]+i \sigma_D[\omega]/\omega.$ 

On the other hand, the case where the number of carriers $n$ and the inverse scattering lifetime $\gamma$ depend on time represents a material in an out-of-equilibrium state. For example, an increase in the number of carriers may describe the transient photodoping of a semiconductor during a femtosecond laser excitation, while its decrease could be due to the subsequent carrier recombination. Similarly, a change in $\gamma$ may describe the increased number of scatterings triggered by the increased phonon temperature after an excitation. 
We write the Drude parameters as the sum of their equilibrium values, $\gamma^{[0]}$ and $n^{[0]}$, and a time-dependent part, $\gamma^{[1]}[t]$ and $n^{[1]}[t]$. Similarly, if we assume that the time variations are small, we can write the fields and the current as a sum of their value they would have if the material were at equilibrium ($E^{[0]},H^{[0]},J^{[0]}$) and their first perturbative order ($E^{[1]},H^{[1]},J^{[1]}$). For a more thorough description, kindly see Eq.~\ref{eq:expansion1} and Eq.~\ref{eq:expansion2} in Appendix.~\ref{A1:expansionMDsystem}.

In this way, the 0-th order fields will satisfy the unperturbed Maxwell-Drude system with constant equilibrium Drude parameters (as shown in Appendix.\ref{A1:expansionMDsystem}, Eq.~\ref{eq_MaxwellDrude0th}), and it can be solved using the standard Transfer Matrix Method.\cite{yang_removal_2021} The 1-st order set of equations will then be expressed as
\begin{equation} \label{eq_Maxell_1st_main}
\begin{split}
	&\partial_z  E^{[1]}\left[z,\omega\right]=- i\,\omega\,\mu[\omega]\,   H^{[1]}\left[z,\omega\right],\\
	&\partial_z  H^{[1]}\left[z,\omega\right]=- i\,\omega\,\epsilon_T[\omega]\, E^{[1]}\left[z,\omega\right] + \mathcal{J}[z,\omega],
\end{split}
\end{equation}
where the source term is given by
\begin{align}\label{sourceterm_main}
	\mathcal{J}[z,\omega]=& \sigma_{D}^{[0]}[\omega] \\
	\times &\mathcal{F}\left[ \frac{n^{[1]}[t]}{n^{[0]}}  E^{[0]}[z,t]  - \frac{m \, \gamma^{[1]}[t]}{n^{[0]}\,e^2}  J^{[0]}[z,t],\omega\right]. \nonumber
\end{align}
and has the dimensionality of a volume current. Here we assume $\gamma^{[0]}$, $\gamma^{[1]}$ $n^{[0]}$, $n^{[1]}$ as known and either they should be provided as inputs (or fitted to experiments) or are assumed to be obtained by other methods.

The 1-st order set of equations in Eqs.~\ref{eq_Maxell_1st_main} cannot be solved using the standard TMM approach. The equations are not homogeneous, and a source term that depends on $E^{[0]}_n$ and $H^{[0]}_n$ is present. 
To solve Eqs.~\ref{eq_Maxell_1st_main} we first construct the general solution. The construction of the general solution is similar to the derivation of the Transfer Matrix with an additional source term developed in Ref~\onlinecite{Yang_transfer_2021} (for more detail, please refer to Appendix.~\ref{A2:TMM_methods}). We count the layers from left to right using an index $n$ starting from 1 up to the number of layers $N$. We assume the multilayered system to be sandwiched by air. The air on the left (right) has index 0 (N+1) and is assumed to be semi-infinite. With this approach, the 1-st order correction to the fields within a given layer $n$ is
\begin{equation} \label{singleLayerFields1st_main}
	\begin{bmatrix} E_n^{[1]}\left[\omega,z\right]\\ H_n^{[1]}\left[\omega,z\right]\end{bmatrix} =\bar{\bar{a}}_n\left[ \omega,z \right]\bar{f}_n\left[ \omega\right] + \sum_l \mathcal{J}_{n}[k_l, \omega]\, \bar{b}[\omega,k_l,z]. 
\end{equation}
where $\bar{f}_n$ is a vector constructed using the right propagating waves amplitude $f_n^>$ and the left propagating waves amplitude $f_n^<$ in the n-th layer.
Using the expression in Eq.~\ref{singleLayerFields1st_main}, the field continuity conditions at all the layers' interfaces can be written. We obtain that the wave intensities in any two layers $n<M<m$  are linked by the expression,
\begin{align} \label{1stTMM2_main}
	&\begin{bmatrix} f^{[1]>}_{m} \\ f^{[1]<}_{m} \end{bmatrix}  = \bar{\bar{T}}_{[n,m]}\begin{bmatrix} f^{[1]>}_{n} \\ f^{[1]<}_{n} \end{bmatrix} +\sum_l \mathcal{J}[k_l, \omega]\, \bar{\bar{T}}_{[M,m]} \\
	& \times \left(\bar{\bar{a}}_N^{-1}\left[d_N \right] \bar{b}[\omega,k_l,d_N]- \bar{\bar{a}}_N^{-1}\left[0 \right] \bar{b}[\omega,k_l,0]\right), \nonumber
\end{align}
where $\bar{\bar{T}}_{[n,m]}$ is the transfer matrix from the n-th layer in the system to the m-th layer, $\bar{\bar{a}}$ is the element matrix of the transfer matrix, and $\bar{b}$ is a coefficient collection vector, the thickness of the layers is denoted by $d_n$, and $k_l$ is the spatial Fourier component (for detailed derivation and forms please refer to Ref.~\onlinecite{yang_removal_2021,Yang_transfer_2021} and the Appendix.~\ref{A2:TMM_methods}).

With the above expression and the final expressions from Ref.~\onlinecite{yang_removal_2021} Eq.~9 for the standard THz transmission and Ref.~\onlinecite{Yang_transfer_2021} Eq.~21  for the THz emission, we can show that the overall THz transmission results up to the first perturbative order can be written into one single expression as:
\begin{align}\label{eq:overall_expression}
    f^{>}_{N+1} = f^{[0]>}_{0}*t_a+(J^{>}_{emit}-t_b J^{<}_{emit})
    +(J^{>}_{pert.}-t_b J^{<}_{pert.}) 
\end{align}
where,
\begin{align}
    t_a &= \frac{{\bar{\bar{T}}_{[0,N+1],11}}{\bar{\bar{T}}_{[0,N+1],22}}-{\bar{\bar{T}}_{[0,N+1],12}}{\bar{\bar{T}}_{[0,N+1],21}}}{{\bar{\bar{T}}_{[0,N+1],22}}},\\
    t_b &= \frac{\bar{\bar{T}}_{[0,N+1],12}}{\bar{\bar{T}}_{[0,N+1],22}},
\end{align}
with the further subscriptions of $\bar{\bar{T}}_{[0,N+1]}$ refer to the matrix elements.

The first term in Eq.~\ref{eq:overall_expression} refers to the THz transmission through the unperturbed sample (the transmission of the THz probe with pump off, $E_{trans}^{off}$), which can be used to describe the THz-TDS results. The second term refers to the THz emission from one single layer ($E_{emit}$), which can be used to describe the STE results. Finally, the third term refers to the correction of the transmission during the OPTP case (the transmission difference of the THz probe through a sample with the pump on and 
pump off, $E_{trans}^{on}-E_{trans}^{off} = \Delta E$). 

Finally, the 0-th order and 1-st order electric fields can be calculated as,
\begin{align}
    E_{N+1}^{[0]} &= f_{N+1}^{[0]>}*\bar{\bar{a}}_{N+1,11}\left[\omega,0\right]\\
    E_{N+1}^{[1]} &= f_{N+1}^{[1]>}*\bar{\bar{a}}_{N+1,11}\left[\omega,0\right]
\end{align}
where the further index for $\bar{\bar{a}}$ is the element of the matrix. (For a more detailed calculation of $f_{N+1}^{[0]>}$ and $f_{N+1}^{[1]>}$ please refer to the Appendix.~\ref{A2:TMM_methods})
A summary of the relationships between the equation and the experiment labels can be seen in Table.~\ref{tab:Expressions}.

\begin{table}[h]
\caption{\label{tab:Expressions} The theoretical expression and the experimental labeling.  }
\begin{ruledtabular}
\begin{tabular}{ccc}
\textrm{Equation}&\textrm{Theoretical}&\textrm{Experimental}\\
\textrm{terms}&\textrm{labels}&\textrm{labels}\\
\colrule
$f^{[0]>}_{0}*t_a$ & $E_{N+1}^{[0]}$ & $E_{trans}^{off}$\\
$J^{>}_{emit}-t_b J^{<}_{emit}$  & -- & $E_{emit}$  \\
$J^{>}_{perb}-t_b J^{<}_{perb}$& $E_{N+1}^{[1]}$ & $E_{trans}^{on}-E_{trans}^{off} = \Delta E$\\
\end{tabular}
\end{ruledtabular}
\end{table}

\section{Results}

To show the capabilities of the framework developed above, we apply the approach to two test cases. Firstly we show how a THz probe traverses a laser excited sample. We choose a quartz(1mm)/Fe(3nm)/Pt(3nm) heterostructure, as an example, but of course the proposed approach can be applied to any stack of materials. Although the chosen heterostructure is a commonly used spintronics THz emitter, in the first test case, we initially ignore the THz that is produced within the sample. The produced THz will be then explicitly treated and its interference with THz probe pulse accounted for in the second test case.

We set the dielectric response of quartz at equilibrium by using a real frequency-independent refractive index of $2.01$ \cite{davies_temperature-dependent_2018,naftaly_terahertz_2007} (yet notice that the model works for any generic frequency-dependent response). Also, we model the dielectric response of the metallic layers using the Drude model and set the plasma and damping frequency to 4.091eV and 0.018eV for Fe and 5.145 eV and 0.069eV for Pt, respectively.\cite{ordal_optical_1985} No further contributions to the response of Fe and Pt at equilibrium have been considered (yet a further contribution to the dielectric function with a generic frequency-dependence can be included). In addition, for simplicity, we considered only the excitation of Pt layer. The excitation of the Fe layer will simply add another term to Eq.~\ref{eq:overall_expression}. If the excitation and de-excitation timescales for both Pt and Fe are similar, their contributions to the THz response will simply sum up in phase. Interesting effects can happen if the dynamics of Fe and Pt do not happen on the same timescales, however such study goes beyond the scope of the present work.

In the following example, we will explicitly address only the case of $\gamma^{[1]}\neq 0$ and $n^{[1]}=0$. We also neglect any effect of the optical laser excitation on quartz. Finally we assume that Pt experiences an increase $\gamma^{[1]} (t)$ in its Drude scattering rate in the form
\begin{equation}   \label{fun:timeprofile1}  
\gamma^{[1]} (t)= h\; \frac{e^{-\frac{t-t_0}{\tau_{\tiny\mbox{decay}}}}}{e^{-\frac{(t-t_0)-\sfrac{\tau_{\tiny\mbox{rise}}}{2}}{\sfrac{\tau_{\tiny\mbox{rise}}}{4}}}+1} ,
\end{equation}
where $h$ controls  the  maximum scattering rate change, $\tau_{\tiny\mbox{rise}}$ is the increase time and $\tau_{\tiny\mbox{decay}}$ is the following decrease time, $t_0$ is the time position of the excitation.
We stress that the focus of this work is to describe the propagation of THz waves through an optically excited system. Eq.~\ref{fun:timeprofile1} is meant purely as an example, mimicking common excitation and thermalization dynamics. However, as already mentioned earlier, a proper form for $\gamma^{[1]}$ and $n^{[1]}$ should be obtained with other methods in future analysis.
In addition, a theoretically built THz probe pulse with a central frequency ($f_c$) of 1.5THz is used in the following calculations.


\subsection{The subpicosecond excitation timescale} 

The OPTP technique is usually used to have information on the de-excitation dynamics of a laser-excited sample \cite{beard_terahertz_2002,baxter_time_2007}. When the de-excitation timescale is longer than the THz probe time-width, the former can be easily extracted from experiments. However, that is not true anymore for the fast excitation timescale. This is usually in the sub-picosecond timescale and, therefore, shorter than the THz pulse time width. At delay values where the THz probe overlaps with the fast excitation of the sample, the temporal dynamics of the OPTP signal are dominated by the probe pulse time-width and are not representative anymore of the excitation timescale. Nonetheless, we will demonstrate that the time-resolved spectra obtained by OPTP experiments contain enough information to evaluate the timescale of the fast excitation after the laser pump, even when it is much shorter than the probe pulse time-width. We remind that, although we are nominally using a Fe/Pt heterostructure as an example, we will not be describing the emission at this stage: therefore what we will find below applies to any fs-laser-excited heterostructure probed by THz pulses. 

\begin{figure}[tb]
\centering
\includegraphics[width=\linewidth]{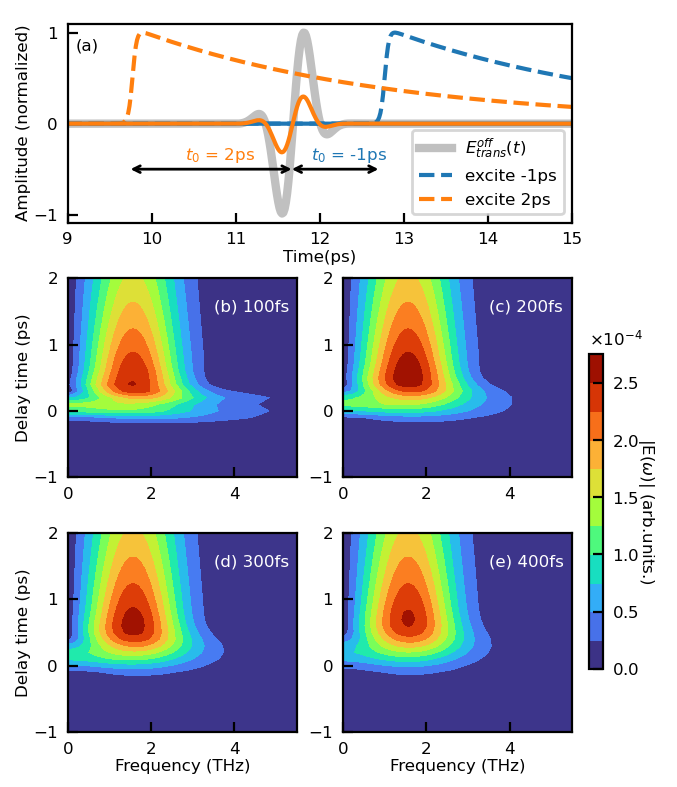}
\caption{(a) Shows the excitation curve $\gamma^{[1]}$ at different initial time delays and its corresponding 1-st order correction term $E_{trans}^{on}-E_{trans}^{off}$ change. (Note that the profiles are normalized and not in scale, it only shows a comparison in the shapes). (b)-(e) Show the frequency maps of $|E_{trans}^{on}-E_{trans}^{off}|$ at different increase times ($\tau_{\tiny\mbox{rise}}=$100fs, 200fs, 300fs, and 400fs) as a function of frequency and time delay change.}
\label{fig:Type1}
\end{figure}

We assume the temporal shape of $\gamma^{[1]} (t)$ known and vary the pump-probe time delay. Fig.~\ref{fig:Type1}(a) shows the respective time positions of the pump and probe pulses, where the time axis is centered around the THz probe pulse and the time dependence of the change in the scattering rate $\gamma^{[1]} (t)$ is shifted in time to simulate the experimental variable delay time. Figs.~\ref{fig:Type1}(b-e) shows the time-resolved THz probe transmission differential spectra (pump-on minus pump-off) for four different $\tau_{\tiny\mbox{rise}}$ times ($\tau_{\tiny\mbox{decay}}$ has been kept the same for all the four cases). 

At sufficiently large negative delay times in Figs.~\ref{fig:Type1}(b-e), the probe pulse arrives well before the laser excitation (dashed blue line in Fig.~\ref{fig:Type1}(a)). The transmitted THz probe is not altered: the difference between the transmitted probe with the pump on and off (full blue line in Fig.~\ref{fig:Type1}(a)) is 0. On the other hand, when the probe pulse arrives well after the sharp excitation dynamics right after the laser excitation (dashed orange line in Fig.~\ref{fig:Type1}(a)) the transmitted THz probe is altered (full orange line in Fig.~\ref{fig:Type1}(a)) mostly in amplitude while its temporal shape, and therefore the spectrum, is not altered. This is evident from Figs.~\ref{fig:Type1}(b-e), where the time-resolved spectra for sufficiently large positive delays simply decreases in amplitude. The time evolution of the differential transmission gives a clear indication of the de-excitation dynamics of the sample $\tau_{\tiny\mbox{decay}}$. On the other hand, the spectrum of the differential transmission does not contain any interesting information since it mostly reproduces the probe pump spectrum. This happens because the change in $\gamma^{[1]} (t)$ is relatively slow, and a quasi-static description of the THz propagation could be adopted.

Conversely the situation is very different when the delay is such that the probe pulse overlaps with the fast rise dynamics in the excitation (small delays in Figs.~\ref{fig:Type1}(b-e)). Here we notice two important features. The timescale over which the differential spectrum shows overlap between the fast timescale $\tau_{\tiny\mbox{rise}}$ and the probe pulse is rather independent on the rise time $\tau_{\tiny\mbox{rise}}$ itself. This is indeed because the overlap time is mostly controlled by the THz pulse time-width. For this reason the temporal duration of this feature does not give any insight into the time evolution of the material's dielectric properties. What is instead evident is that the spectrum during the overlap is strongly affected by the $\tau_{\tiny\mbox{rise}}$ time (see Figs.~\ref{fig:Type1}(b-e)). 

The analysis of the spectrum map reveals two interesting findings. Firstly, it is observed that the spectrum is the broadest when the excitation overlap with the central time position of the probe. Secondly, a faster increase time for the excitation profile results in a broader spectrum map. We find that it is possible to obtain a fairly accurate estimation of the fast rise time $\tau_{\tiny\mbox{rise}}$. This can be achieved by selecting the delay time at which the spectrum is the widest. At that time delay one should identify the frequency$f_{1\%}$  at which the spectrum has 1\% of its maximum amplitude. The fast rise time $\tau_{\tiny\mbox{rise}}$ can then be estimated using
\begin{equation} \label{eq:estimation}
    \tau_{\tiny\mbox{rise}} \approx \frac{1}{f_{1\%}-2f_{c}},
\end{equation}
where $f_c$ is the central frequency of the THz probe.

\begin{figure}[tb]
\centering
\includegraphics[width=\linewidth]{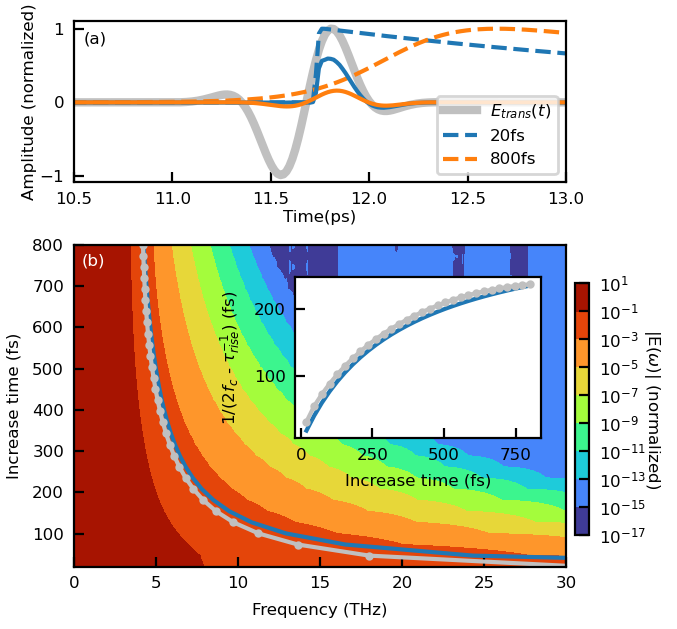}
\caption{(a) Shows the excitation curve $\gamma^{[1]}$ at a fixed time $t_0$ but with different increase time parameters $\tau_{\tiny\mbox{rise}}  = 20fs$ and $\tau_{\tiny\mbox{rise}}  = 800fs$ (Note that the profiles are normalized and not in scale, it only shows a comparison in the shapes). (b) Normalized frequency map of OPTP signal as a function of the frequency and increase time $\tau_{\tiny\mbox{rise}} $.
The pump-probe delay time is fixed at $t_0 = 0 ps$. The gray dotted line shows the region where the spectrum amplitude is 1\% of its maximum ($f_{1\%}$). The blue line is $f_c*2+1/\tau_{\tiny\mbox{rise}} $. Inset: The inverse plotting of the extracted gray dotted line and blue line on the map. }
\label{fig:Type2}
\end{figure}

To prove that the formula above provides a good estimation of the fast rise time $\tau_{\tiny\mbox{rise}}$ even when it is much shorter than the THz probe time-width, we fix the delay time between the probe and excitation at the point of overlap and compute the differential spectrum for varying rise times $\tau_{\tiny\mbox{rise}}$ (as shown in Fig.~\ref{fig:Type2}(a), where the reader should notice the logarithmic intensity scale). The blue line in the main figure of Fig.~\ref{fig:Type2}(b) represents simply the conversion of the $\tau_{\tiny\mbox{rise}}$ to frequency. It is already evident how the inverse rise time remains approximately parallel to the level lines. After some fine tuning we find that the best estimation of $\tau_{\tiny\mbox{rise}}$ is obtained by Eq.~\ref{eq:estimation} where a correction including the central frequency of the probe has been added. The error occurred when using the estimation above is shown in the inset in Fig.~\ref{fig:Type2}(b).

The above-mentioned correspondence, although very convenient due to its simplicity, is not perfect. For an accurate extraction of the excitation time, a comparison between experiments and theory is needed. It is however important to note how the method can clearly resolve timescales (hundreds of fs) that are shorter than the period and the time width of the employed radiation ($\approx$ 2ps in the presented example).

The analysis above can be conducted easily on any excited multilayer that does not produce THz upon laser excitation. It is however possible to perform the above analysis in heterostructures used for spintronics THz emitters.  In this case, the time-resolved spectra cannot be simply constructed by subtracting the THz probe transmission with pump on by the THz probe transmission without pump: the THz radiation internally generated by the sample must be subtracted as well. This can be achieved because THz signal produced by the inverse-spin Hall effect is proportional to the optical pump intensity, but not to the THz probe, while the spectral maps in Fig.\ref{fig:Type1} are proportional to both the the optical pump intensity and the THz probe amplitude.

\subsection{Pump-THz emission time delay}

\begin{figure*}[tb]
\centering
\includegraphics[width=\linewidth]{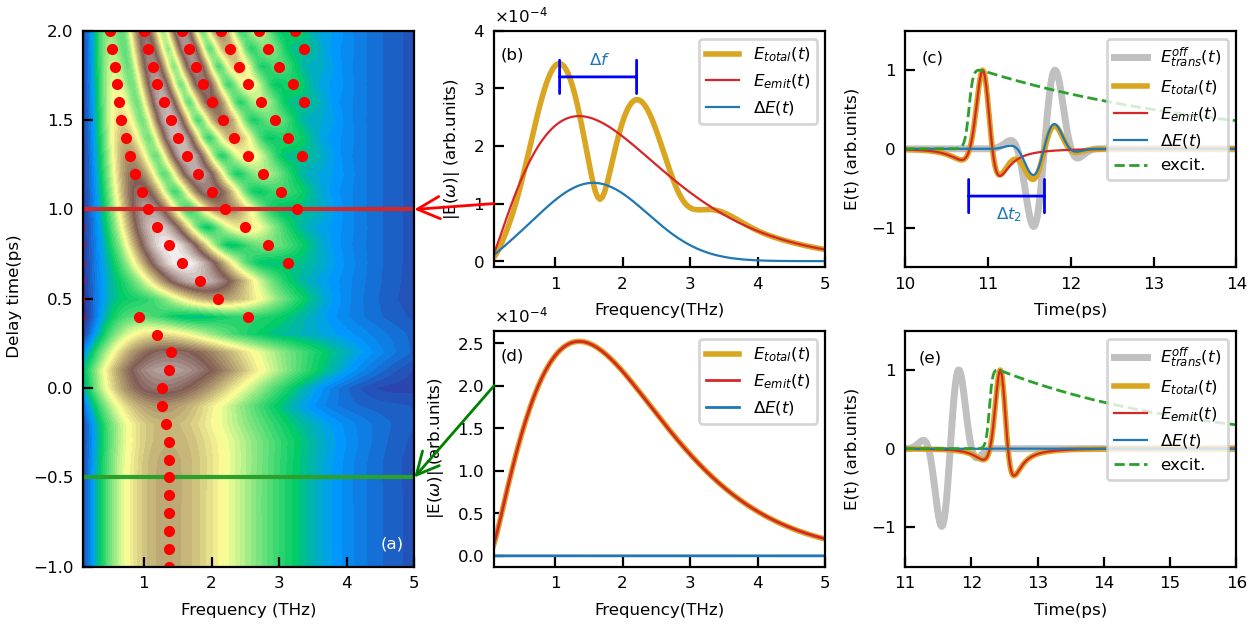}
\caption{Pump-Probe analysis of Spintronic THz emitter:(a) shows a overall spectrum map ($E_{emit}+ \Delta E$) as a function of frequency and delay time. The red dots on the map shows the peaks of the spectrum at different time delays. (b)-(c) refers to the spectrum and its corresponding THz profiles in the time domain for case 1 (probe sent in after the excitation). (d)-(e) refers to the spectrum and its corresponding THz profiles in the time domain for case 2 (probe sent in before the excitation). Notice that the profiles in the time domain are normalized. }
\label{fig:Type3}
\end{figure*}

We here focus completely on the analysis of spintronics THz emitters (STE): these are (in their simplest configuration) a bilayer system with a ferromagnetic (FM) layer and a nonferromagnetic (HM) layer, where after excitation, a spin current transfers from FM to HM, undergoes ISHE to become a charge current, and eventually emit terahertz radiation.\cite{kampfrath2013terahertz,seifert_efficient_2016}
After determining the exact time when the excitation and the probe overlap in the system, we are now able to proceed to another type of spectrum analysis, which is when the THz emission and 1-st order correction ($\Delta E$) exist at the same time, and their interference patterns can be analyzed to extract further information.

In this analysis, we obtained an overall spectrum map of changing time delay between the excitation and probe, only that now the THz emission has not been subtracted (as shown in Fig.~\ref{fig:Type3}). The THz emission will happen after a certain time ($\Delta t$ shown in Fig.~\ref{fig:OPTP}(b)) of the excitation, which is usually believed to be only tens of fs\cite{wang_ultrafast_2018}. Notice that the shape of this interference spectrum depends on the relative amplitude of the emitter THz and the 1-st order correction to the transmission of the THz probe. Maximal interference (and, therefore, the easiest spectral map to analyze) is obtained when the two contributions have similar amplitudes. This can be obtained experimentally by varying the probe intensity. 

In the spectrum map we observe that when we send the probe in after the excitation, more than one peak will exist in the overall spectrum (red dots in Fig.~\ref{fig:Type3}(a)). We now focus on two specific cases: when the probe is sent in after the system is excited (Fig.~\ref{fig:Type3}(b)-(c)) and when the probe is sent in before the system is excited(Fig.~\ref{fig:Type3}(d)-(e)). We find that the spectrum shows interference peaks for case 1, while for case 2, the overall spectrum is equal to the THz emission spectrum. 

Interestingly, the frequency difference between the peaks ($\Delta f$) of the overall spectrum for case 1 is going to give us valuable information about the time difference between the emission pulse and the correction pulse ($\Delta t_2$ in Fig.~\ref{fig:OPTP} and Fig.~\ref{fig:Type3}(c)). In fact, we see that
\begin{equation}
    \Delta t_2 \approx \frac{1}{\Delta f}.
\end{equation}

\section{Conclusions}

In this study we have provided a robust framework for the description of THz radiation propagating through optically excited systems beyond the quasi-static approximation. More specifically, our approach can be used to describe the spectral changes in the transmitted THz probe when it traverses a multilayer undergoing ultra-fast changes in its dielectric properties. We have shown that OPTP spectral maps provide a large amount of information and shown how to extract, for instance, the ultrafast excitation time constant, even when it is shorter than the THz probe time-width.

Our approach can also be extended to incorporate the THz emitted by spintronics THz emitters. We have then shown how.

We have shown that by incorporating the results from the analyses above, our study can provide important insights into the temporal dynamics of ultrafast laser excitations via OPTP experiments. Furthermore, we have shown how, by using the analyses above, we were able to determine two crucial time delays: the time delay between the probe and the excitation ( $\Delta t_1 $ in Fig.~\ref{fig:OPTP}) using the spectrum map of changing time delay for analysis type 1, and the time difference between the THz emission and the probe pulse ($\Delta t_2 $ in Fig.~\ref{fig:OPTP}) using the overall spectrum map of THz emission. By combining these results, we can determine the time difference between the THz emission and the excitation ($\Delta t = \Delta t_1 - \Delta t_2$), which is a critical parameter for the accurate interpretation of THz emission experiments.

\bibliography{sample}

\appendix
\section{Perturbation expansion of the Maxwell-Drude system}\label{A1:expansionMDsystem}

In this section, we provide more details on the Perturbation expansion of the Maxwell-Drude system. We account for fs-laser driven ultrafast dynamics by assuming that the Drude parameters in Eq.~\ref{eq_MaxwellDrude_main} in Sec.~\ref{Theoretical_background} are a combination of their equilibrium values, $\gamma^{[0]}$ and $n^{[0]}$, and a time-dependent component, $\gamma^{[1]}$ and $n^{[1]}$, as outlined in the main text:
\begin{align}
	\gamma[z,t]&=\gamma^{[0]}+  \gamma^{[1]}[z,t],\\
	n [z,t] &= n^{[0]} + n^{[1]}[z,t].\label{eq:expansion1}
\end{align}
Please note that we also consider the possibility of position-dependent variations. However, for conciseness, we will omit explicit position dependence throughout the remaining part of this section, as it doesn't affect the analysis. As mentioned in the main text, assuming small time variations, we can express the fields and current up to the first perturbative order as:
\begin{equation} \label{eq:expansion2}
\begin{split}
	E[z,t]&\approx E^{[0]}[z,t] + E^{[1]}[z,t],\\
	H[z,t]&\approx H^{[0]}[z,t] + H^{[1]}[z,t],\\
	J[z,t]&\approx J^{[0]}[z,t] +  J^{[1]}[z,t],
\end{split}
\end{equation}
where the two orders must satisfy two different sets of equations. In this way, the set of equations at the 0-th order adheres to the unperturbed Maxwell-Drude system as presented in Eqs.~\ref{eq_MaxwellDrude_main}, featuring constant equilibrium Drude parameters. This set of equations is expressed as follows:
\begin{align} \label{eq_MaxwellDrude0th}
    &\partial_{z}  E^{[0]}[z, \omega]=-i \omega \mu[\omega]  H^{[0]}[z, \omega] \nonumber,  \\
    &\partial_{z}  H^{[0]}[z, \omega]=-i \omega \epsilon_T[\omega] E^{[0]}[z, \omega],
\end{align}
This set of equations can be solved analytically using a standard Transfer Matrix Method approach \cite{yang_removal_2021}. Then, the 1-st order set of equations will read,
\begin{align} \label{eq_MaxwellDrude1st}
    \partial_{z}  E^{[1]}[z, \omega]=&-i \omega \mu[\omega]  H^{[1]}[z, \omega] \nonumber,  \\
    \partial_{z}  H^{[1]}[z, \omega]=&-i \omega \epsilon_B[\omega] E^{[1]}[z, \omega]+ J^{[1]}[z, \omega] , \\
    \partial_{t}  J^{[1]}[z, t]=&-\gamma^{[0]} J^{[1]}[z,t]+\frac{n^{[0]} e^{2}}{m}  E^{[1]}[z, t]- \nonumber\\
    &   \gamma^{[1]} J^{[0]}[z,t]+\frac{n^{[1]} e^{2}}{m}  E^{[0]}[z, t].  \nonumber
\end{align}
In relation to this first-order set of equations, we perform the Fourier Transforms on the final equation of Eqs. \ref{eq_MaxwellDrude1st}. This transformation allows us to simplify the system into the Maxwell equations, as demonstrated in the main text from Eq. \ref{eq_Maxell_1st_main} to Eq. \ref{sourceterm_main}.

For ease of reference, we subsequently represent Eq. \ref{sourceterm_main} using its spatial Fourier transform:
\begin{equation}
	\mathcal{J}[z,\omega]=\sum_l \mathcal{J}[k_l, \omega] \exp[i\, k_l\, z],
\end{equation}
It's important to highlight that $k_l$ represents the spatial Fourier wavevector of the source term's spatial distribution, distinct from the light wavevector. Thus, in general, $k_l \neq \omega \sqrt{\epsilon \mu}$. To maintain conciseness, we will omit the Fourier summation in the subsequent formulas and reintroduce it solely in the final expression.

\section{Perturbative Transfer Matrix Method}\label{A2:TMM_methods}

Before addressing the Perturbative Transfer Matrix Method (PTMM), our discussion commences with a brief revisitation of the traditional Transfer Matrix Method (TMM) and its expansion into the Transfer Matrix Method with Source (TMMS). Detailed derivations employing symbols consistent with this article can be found in Ref.~\onlinecite{yang_removal_2021} and Ref.~\onlinecite{Yang_transfer_2021}, but the fundamental results of TMM and TMMS are outlined below for the reader's convenience.

For the analysis of electromagnetic radiation propagation within a multilayered heterostructure under normal incidence, we adopt specific notation. The layers are sequentially numbered from 1 to N, oriented left to right, with an index denoted as $n$. The heterostructure is encapsulated by air layers, indexed as 0 on the left and N+1 on the right, both extending infinitely. The thickness of each layer is symbolized by $d_n$. Furthermore, we work under the assumption that negligible current accumulates between layers, ensuring the uninterrupted continuity of electric and magnetic fields across interfaces.

\subsection{TMM: 0-th order propagation}\label{subsec:0thOrderPro}

At the 0-th order, we adhere to the standard TMM approach. Within a specific layer indexed as $n$, the 0-th order electric and magnetic fields can be represented as the product of a $2\times 2$ matrix and a vector containing the amplitudes of the rightward propagating waves $f^{[0]>}_n$ and leftward propagating waves $f^{[0]<}_n$:
\begin{equation}\label{0thTMM1}
\begin{bmatrix} E_n^{[0]}\left[\omega,z\right]\\ H_n^{[0]}\left[\omega,z\right]\end{bmatrix} =  \bar{\bar{a}}_n\left[\omega,z\right]
\begin{bmatrix} f^{[0]>}_n\left[ \omega\right] \\ f^{[0]<}_n\left[ \omega\right] \end{bmatrix}, 
\end{equation}
where
\begin{align}
    &\bar{\bar{a}}_n\left[\omega,z\right]=\begin{bmatrix} e^{i \omega \sqrt{\epsilon_n \mu_n} z } &e^{-i \omega \sqrt{\epsilon_n \mu_n} z} \\ 
    -\sqrt{\frac{\epsilon_n}{\mu_n}}e^{i \omega \sqrt{\epsilon_n \mu_n} z }&{\sqrt{\frac{\epsilon_n}{\mu_n}} e^{-i \omega \sqrt{\epsilon_n \mu_n} z}}\end{bmatrix}.
\end{align}

By imposing the necessary continuity equations for the fields at the interfaces between layers, one deduces a relationship linking the field amplitudes in the air layers situated on the left and right sides of the sample:
\begin{equation}\label{0thTMM2}
	\begin{bmatrix} f^{[0]>}_{N+1} \\ f^{[0]<}_{N+1} \end{bmatrix}  = \bar{\bar{T}}_{[0,N+1]}\begin{bmatrix} f^{[0]>}_{0} \\ f^{[0]<}_{0} \end{bmatrix},
\end{equation}
where $\bar{\bar{T}}_{[0,N+1]}$ is the Transfer Matrix given by,
\begin{equation}\label{eq:T_0N}
    \bar{\bar{T}}_{[0,N+1]}=\bar{\bar{a}}_{N+1}^{-1}\left[0 \right] \left(\prod_{j=N}^{1} \bar{\bar{a}}_j[d_j] \bar{\bar{a}}_j^{-1}[0] \right)\bar{\bar{a}}_0\left[0 \right].
\end{equation}
Furthermore, when contemplating the transfer matrix between any two layers (indexed as $n$ and $m$, with $n < m$) within the system, it can be formulated in a more general manner:
\begin{equation}\label{eq:T_nm}
    \bar{\bar{T}}_{[n,m]}=\bar{\bar{a}}_{m}^{-1}\left[0 \right] \left(\prod_{j=m-1}^{n+1} \bar{\bar{a}}_j[d_j] \bar{\bar{a}}_j^{-1}[0] \right)\bar{\bar{a}}_n\left[d_n \right]. 
\end{equation}

By solving the set of linear equations presented in Eq.~\ref{0thTMM2} and utilizing the field expressions from Eq.~\ref{0thTMM1}, we can compute the 0-th order $E^{[0]}_n$ and $H^{[0]}_n$ fields within any layer.

\subsection{TMMS and PTMM: 1-st order propagation}\label{subsec:1stOrderPro}

The set of equations at the 1-st order, as given in Eqs.~\ref{eq_MaxwellDrude1st}, cannot be resolved through the conventional TMM. These equations are non-homogeneous and incorporate a source term that hinges on $E^{[0]}_n$ and $H^{[0]}_n$. To address Eqs.~\ref{eq_MaxwellDrude1st}, we initiate by establishing a comprehensive solution, akin to the approach undertaken in Ref.~\onlinecite{Yang_transfer_2021}.

The solution can be formulated by combining the general solution related to the homogeneous system with a particular solution. The general solution of the corresponding homogeneous system is achieved by:
\begin{equation} \label{eq_associated_homog}
\begin{split}
	&\partial_z  E^{[1]}\left[z,\omega\right]=- i\,\omega\,\mu[\omega]\,   H^{[1]}\left[z,\omega\right], \\
	&\partial_z  H^{[1]}\left[z,\omega\right]=- i\,\omega\,\epsilon_T[\omega]\,   E^{[1]}\left[z,\omega\right] ,
\end{split}
\end{equation}
This can be readily constructed using the standard TMM. We look for a particular solution in the form, 
\begin{align}
	E^{[1]}\left[z,t\right]=& E e^{i(k_l z-\omega t)}, \\
	H^{[1]}\left[z,t\right]=& H e^{i(k_l z-\omega t)}.
\end{align}
We substitute the above in Eqs.~\ref{eq_MaxwellDrude1st} and obtain the amplitudes of the fields for the particular solution,
\begin{align}
	k_l E=& \mu \omega H,\\
	i k_l H =&i \epsilon \omega E + \mathcal{J}[k_l,\omega].
\end{align}
The particular solution reads,
\begin{equation}\label{sourcefiled}
	\bar{F}\left[\omega,z\right] = 
	\begin{bmatrix} E\left[\omega,z\right]\\ H\left[\omega,z\right]\end{bmatrix} =  \mathcal{J}[k_l,\omega] \;\bar{b}[\omega,k_l,z],
\end{equation}
with 
\begin{equation}
	\bar{b}[\omega,k,z]=\frac{i \,e^{i kz}}{\epsilon \mu \omega^2-k^2}  \begin{bmatrix} \omega \mu\\ k\end{bmatrix} .
\end{equation}

With this approach, the 1-st order correction to the fields within a given layer $n$ can be expressed as the Eq.~\ref{singleLayerFields1st_main}, and the wave intensities can be expressed as Eq.~\ref{1stTMM2_main}.

Finally, considering the two semi-infinite air layers, Eq.~\ref{1stTMM2_main} can be recast as,
\begin{align} \label{1stTMM_simplified1}
	&\begin{bmatrix} f^{[1]>}_{N+1} \\ f^{[1]<}_{N+1} \end{bmatrix}  = \bar{\bar{T}}_{[0,N+1]}\begin{bmatrix} f^{[1]>}_{0} \\ f^{[1]<}_{0} \end{bmatrix}+ \begin{bmatrix} J^{>}_{M} \\ J^{<}_{M} \end{bmatrix}.
\end{align}
The equation given as Eq.~\ref{1stTMM_simplified1} consists of a pair of linear, frequency-dependent equations.
For any fixed source term and for a known incoming field, Eq.~\ref{1stTMM_simplified1} can be applied to compute the transmitted and reflected waves in any layer.

When accounting for both the emission source (THz emission from spintronic THz emitter) and the correction source (THz transmission change from OPTP experiments) concurrently, the comprehensive expression becomes:
\begin{align} \label{1stTMM_simplified2}
	&\begin{bmatrix} f^{[1]>}_{N+1} \\ f^{[1]<}_{N+1} \end{bmatrix}  = \bar{\bar{T}}_{[0,N+1]}\begin{bmatrix} f^{[1]>}_{0} \\ f^{[1]<}_{0} \end{bmatrix}+ \begin{bmatrix} J^{>}_{emit} \\ J^{<}_{emit} \end{bmatrix}+ \begin{bmatrix} J^{>}_{perb} \\ J^{<}_{perb} \end{bmatrix}.
\end{align}

\subsection{Reflection and transmission}

We take $f^{[0]>}_{0}(\omega)$ as the incoming THz probe (from the left), and we assign $f^{[0]<}_{N+1}(\omega)$ as $0$, given that no wave is incoming from the right. By solving Eq.~\ref{0thTMM2} for $f^{[0]>}_{N+1}$ and $f^{[0]<}_{0}$, the profiles of the 0-th order transmitted and reflected waves are determined, as shown in Appendix.~\ref{subsec:0thOrderPro}.

Similarly, when solving Eq.~\ref{1stTMM_simplified1}, we assign $f^{[1]>}_{0}$ and $f^{[1]<}_{N+1}$ as 0. This is because the optical pump induces no alteration in the incoming THz probe and no radiation emanates from the right side.

Upon solving Eq.~\ref{1stTMM_simplified1} for $f^{[1]>}_{N+1}$ and $f^{[1]<}_{0}$, the first-order correction to the transmitted and reflected waves can be deduced, respectively. Notably, the correction to the transmission is:
\begin{equation}
    f^{[1]>}_{N+1} = J^{>}_{M}-\frac{\bar{\bar{T}}_{[0,N+1],12}}{\bar{\bar{T}}_{[0,N+1],22}}J^{<}_{M},
\end{equation}
where the further subscriptions of $\bar{\bar{T}}_{[0,N+1]}$ refer to the matrix elements. Similarly, if both the THz emission source and the correction source are considered, then the equation can be further written into
\begin{align}
    f^{[1]>}_{N+1} = &J^{>}_{emit}-\frac{\bar{\bar{T}}_{[0,N+1],12}}{\bar{\bar{T}}_{[0,N+1],22}}J^{<}_{emit}+\nonumber\\&J^{>}_{perb}-\frac{\bar{\bar{T}}_{[0,N+1],12}}{\bar{\bar{T}}_{[0,N+1],22}}J^{<}_{perb}.
\end{align}
And if the THz transmission shown in Eq.~\ref{0thTMM2} is considered, the final simplified results, which contains the transmitted THz, the emitted THz, and the correction of the THz transmission from the OPTP, can be expressed as Eq.~\ref{eq:overall_expression} in the main text.

\section{Inputs of PTMM}\label{Inputs}
A summary of the inputs of calculation of the PTMM are listed in the table below.

\begin{table}[H]
\caption{Summary of initial inputs.}
\begin{tabular}{|c|c|l|}
\hline
\multicolumn{1}{|l|}{}                                                                            & Inputs                                                                                         & \multicolumn{1}{c|}{Physical meaning}                                                                                                                                                                                                                                                                                                                                                                                                                                                \\ \hline
Geometry                                                                                          & $d_n$                                                                                          & Thickness of each layer                                                                                                                                                                                                                                                                                                                                                                                                                                                              \\ \hline
                                                                                                  & \begin{tabular}[c]{@{}c@{}}$\gamma^{[0]}$ \\ and \\ $n^{[0]}$\end{tabular}                     & {\color[HTML]{333333} \begin{tabular}[c]{@{}l@{}}Equilibrium Drude Parameters \\ for each layer (layer index suppressed)\end{tabular}}                                                                                                                                                                                                                                                                                                                                               \\ \cline{2-3} 
\multirow{-2}{*}{\begin{tabular}[c]{@{}c@{}}Equilibrium \\ materials' \\ properties\end{tabular}} & $\epsilon_B(\omega)$                                                                           & \begin{tabular}[c]{@{}l@{}}Remaining contribution to the \\ equilibrium dielectric response \\ of the material for each layer \\ (layer index suppressed)\end{tabular}                                                                                                                                                                                                                                                                                                               \\ \hline
\begin{tabular}[c]{@{}c@{}}Modification \\ of \\ materials' \\ properties\end{tabular}            & \begin{tabular}[c]{@{}c@{}}$\gamma^{[1]}(t,z)$ \\ and \\ $n^{[1]}(t,z)$\end{tabular}           & \begin{tabular}[c]{@{}l@{}}Time- and position-dependent \\ variations of the Drude parameters \\ (layer index suppressed). \\ These are the consequence of the \\ optical laser pump, electrons \\ thermalization, carrier recombination, \\ cooling through phonons and/or \\ heat transfer to the substrate, etc. \\ We do not address these dynamics in \\ this work and we suppose that the \\ induced changes in the Drude \\ parameters are known by other means.\end{tabular} \\ \hline
                                                                                                  & $f_0^{[0]>}$                                                                                   & \begin{tabular}[c]{@{}l@{}}We assume that the THz probe pulse \\ impinges on the sample from the left. \\ Therefore f {[}0{]}\textgreater{}0 represents the \\ incoming THz pulse E-field profile, \\ which is supposed known.\end{tabular}                                                                                                                                                                                                                                          \\ \cline{2-3} 
\multirow{-2}{*}{THz probe}                                                                       & \begin{tabular}[c]{@{}c@{}}$f_{N+1}^{[0]<}$,\\ $f_{0}^{[1]>}$,\\ $f_{N+1}^{[1]<}$\end{tabular} & These are all 0.                                                                                                                                                                                                                                                                                                                                                                                                                                                                     \\ \hline
\end{tabular}
\end{table}

\end{document}